\documentclass{JAC2000}               

\usepackage{graphicx}

\begin{document}
\title{\flushright{WEAP025}\\[15pt] \centering 
INTRODUCTION OF MODERN SUBSYSTEMS AT THE KEK INJECTOR-LINAC}

\author{N.Kamikubota, K.Furukawa, KEK, Tsukuba, Japan\\
    S.Kusano, T.Obata, Mitsubishi Electric System and Service Co. Ltd., Tsukuba, Japan
}

\maketitle

\begin{abstract}
As an accelerator control system survives over several years, it is often the 
case that new subsystems are introduced into the original control system. 
The control system for the KEK electron/positron injector-linac has been 
using Unix workstations and VME computers since 1993. 
During the eight-year operation, we  
extended the system by introducing 
a) Windows PCs, 
b) PLC controllers with a network interface, and 
c) web servers based on modern information technology. 
Although such new subsystems are essential to improve control 
functionalities,  they often cause communication problems with the 
original control system. 
We discuss the experienced problems, and present our 
solutions for them. 

\end{abstract}

%
%
%
%
\section{INTRODUCTION}

The KEK linac was constructed 
as an injector of the Photon Factory storage ring about 20 years ago \cite{nim80}. 
The first beam of 2.5-GeV electrons was provided in 1982. 
This linac now provides electron/positron beams 
to several rings \cite{pac99-comm}: 
a) 3.5-GeV positrons to the KEKB LER (KEK B-factory Low-energy ring), 
b) 8-GeV electrons to the KEKB HER (High-energy ring), 
c) 2.5-GeV electrons to the PF ring, and 
d) 2.5-GeV electrons to the PF-AR ring. 
The first control system, 
which consisted of mini-computers and CAMAC interfaces \cite{nim86}, 
was replaced by the present control system in 1993 \cite{nim93}. 
The present system 
comprises Unix workstations and VME  computers. 
It has been upgraded occasionally \cite{icale95} 
and used over the past eight years. 

In this article,
we discuss newly introduced subsystems 
in recent years. They were introduced 
in order to improve the control functionalities, and/or 
to enable better maintenance capabilities.  
Three subsystems are described in detail in Section~\ref{Subsystems}. 
The experienced problems between the subsystems and the original control system, 
and their solutions are discussed in Section~\ref{Discussion}.

%
%
\section{New Subsystems}
\label{Subsystems}

\subsection{Control System Overview}

\begin{figure}[htb]
\centering
\includegraphics*[width=82mm]{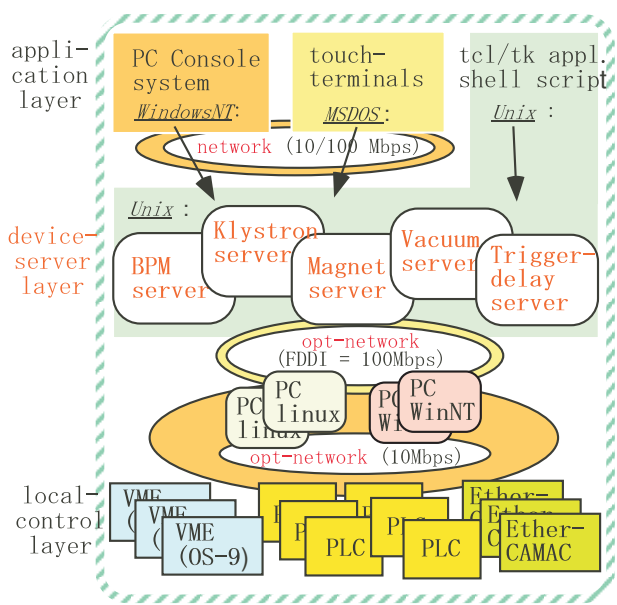}
\caption{Simplified view of the control system.}
\label{overview}
\end{figure}

The present control system comprises 4--6 UNIX workstations, 
27 VME computers with the OS-9 operating system, 
140 PLC (Programmable logic controller) controllers, and 
11 CAMAC interfaces with a network port. 
A home-made RPC (remote procedure call), 
based on TCP/UDP protocols, 
are used for communication between them. 
A simplified view of the control system is shown in 
Fig.~\ref{overview}. 

The total number of control signals is about 6000 (in 2-byte unit). 
Since 1998, when the CP violation study started with the KEKB rings,
the operation of the linac exceeded 7000 hours per year. 
The number of control transactions handled by the control system 
increased every year, and reached 350 transactions per sec. 
in June, 2001 \cite{apac01}. 


\subsection{Windows PC}

Since the end of the 1980'es, we have had strong interest in using PCs.  
The PC-based operator's console system 
started with DOS PCs \cite{nim90}, and later 
reinforced with Windows PCs \cite{opeconsole99}, 
has been successfully used 
for more than ten years. 
In the early phase, 
the following points were preferable for us: 
a) enhanced capability of 2-byte code (Japanese characters) handling, 
b) good development environment of graphic applications, 
and c) low cost. 

The present console system 
comprises about ten PCs (Windows~NT and Windows~2000). 
The surveillance applications for accelerator devices 
were developed in Visual~Basic, and have been used in daily operation. 
The operation log-book using MS-SQL and Access \cite{elog99} 
is extensively used everyday with this console system. 

Communication with the control system, which runs at the Unix workstations, 
is made by a gateway (a Windows PC). 
When the gateway receives a control request from a console PC by the OLE, 
it communicates with the appropriate device server(s) 
by using the RPC protocol (see Fig.~\ref{overview}). 
The gateway and the present console system have been 
successfully used over the past six years.

\subsection{PLC}

The main part of the control system for the KEK linac was renewed 
in 1993 \cite{nim93}. 
However, 
the local controllers (shown as SBC\footnote{Such old local controllers 
were controlled by Single Board Computers with micro-processors \cite{nim86}.} 
in Table~\ref{frontend}) 
remained. 
In recent years the maintenance of these local controllers 
has become difficult. 
Thus, we decided to replace them with new controllers. 

\vspace{-3 mm}
\small
\begin{table}[htb]
\begin{center}
\caption{Replacement of local controllers.}
\begin{tabular}{|l|l|l|l|}
\hline
device and    & before & transition & present \\ 
transition & 1993   & phase & status \\ \hline \hline
Klystron  &CAMAC & VME        & Ethernet \\
'97-'98 & --\textgreater SBC & --\textgreater SBC & --\textgreater PLCx70 \\ \hline
Magnet & CAMAC & VME       & Ethernet \\
'96-'00 & --\textgreater SBC & --\textgreater SBC &--\textgreater PLCx51 \\ \hline
Vacuum & CAMAC & VME       & Ethernet  \\
'96-'97 & --\textgreater SBC & --\textgreater SBC & --\textgreater PLCx18 \\ \hline
Trigger& CAMAC & VME       & Ethernet  \\
'97-now & --\textgreater SBC & --\textgreater SBC & --\textgreater CAMACx11 \\ \hline
BPM    & none  &           & Ethernet \\ 
since '97  &       &           & --\textgreater VMEx19  \\ \hline
\end{tabular}
\label{frontend}
\end{center}
\end{table}
\normalsize
\vspace{-3 mm}

A typical local controller 
should have a) a few hundred I/O points, b) simple but programmable control logic, 
and c) a communication path to the main control system. 
Among some candidates, 
a PLC with a direct network port (Yokogawa FA-M3) was chosen 
for klystron modulators, magnet power-supplies, and vacuum controllers. 
The VME computer was a candidate, but was not selected, because 
the PLC is less expensive.  
The replacements of local controllers since 1996 are summarized in Table~\ref{frontend}. 



It is interesting that 
all of the new controllers shown in Table~\ref{frontend} have an Ethernet port. 
A background fact is that 
world-standard field networks (CAN-bus, Profi-bus, MIL1553, etc.) are not 
popular in Japan, 
and we want to use Ethernet as a field network. 
The use of a standard Ethernet is preferable for long-term maintenance and 
cost reduction. 
In addition, we use optic-fiber cables for the network to local controllers  
in order to avoid electro-magnetic noise from the klystron modulators.

\subsection{Web and Related Topics}

We have 
recently experienced fast improvements of 
information technologies. 
The world-wide-web services at the KEK linac 
started in May, 1994  \cite{lin-web}. 
Up to now, we have developed many web pages 
to inform about the linac 
operation 
status. 

\subsubsection{a) Status of accelerator devices}
The web-server machine is a part of the control system. 
Thus, by using the CGI (Common Gateway Interface) script, 
it is easy to develop a homepage 
to show the 
status of any linac device. 
A large number of pages have already been developed. 

\subsubsection{b) Real-time display by Java and CORBA}
Feasibility studies of a web-based real-time display using Java and CORBA  
have been carried out. 
The measured round-trip time 
between a Java applet and a CORBA server (at an Unix workstation) was  
50 ms \cite{pcapackusa99,icalekusa99}. 
The server 
does not consume CPU resources 
compared with the CGI-based services. 
Recent updates have enabled realistic demonstration of  
the beam-current history at the KEK linac \cite{lin-beamweb}. 

%

%
%
\section{Discussion}
\label{Discussion}

\subsection{Problems with Subsystems}

\subsubsection{a) Windows PC}

The main language for the Windows-based console PCs is 
Visual Basic, while the sources at the Unix side 
have been developed in C language. 
For example, the sources for the RPC use socket (Winsock) functions 
at the Unix (Windows) side. 
Thus, 
the maintenances have been made independently. 
This fact implies that 
when we have some improvements at the Unix side, 
it always takes time for the improvements influence the Windows side. 
 
We have operated a TCP/IP network system which contains 
both Unix workstations and Windows PCs. 
As the number of Windows PCs has increased, 
we experienced 
communication errors by two specific intervals (2 hours and 12 minutes). 
They were removed by changing the default settings of MS Office and 
Samba \cite{pcapac00}. 
We also experienced an accident in which the network burst from Windows PCs 
occupied the network system,  
followed by a short-time mistake of network cable connections.  
At the time of the accident, 
the burst stopped all network modules of the PLC controllers. 
We modified the parameters at the network routers 
so as not to enhance the burst broadcasts.

\subsubsection{b) PLC}

After the KEKB commissioning started in 1997, 
we developed various slow-feedback applications \cite{linfb00} 
in order to realize stable beam injections to the KEKB rings. 
By the end of 1998, 
the CPU capabilities of Unix workstations 
were found to be insufficient for increasing demands. 
The analysis showed that 
the klystron server (see Fig.~\ref{overview}) 
consumed a very large fraction of the CPU resources 
for network communication with PLC controllers. 
In the summer of 1999, 
we prepared on-memory cache areas  
to keep klystron data at the Unix workstations. 
Two linux PCs have been used 
to update the cached data by polling the PLC controllers \cite{pcapac00,icalekusa01}. 
The network traffic decreased to one fourth, and  the problem disappeared. 

Up to now 
we have introduced the cached scheme even for other devices.\footnote{The vacuum 
system started to use cache 
in June, 2000. For magnet power-supplies, we had 
a Windows PC as a gateway since 1997 \cite{pcapac00}. 
The BPM servers use cache from the start of the service in 1997.}. 
Considering the number of control transactions  \cite{apac01}, 
we can conclude that 
the intelligence of the PLCs was not sufficient  for our case. 
Thus, 
we added more intelligent devices (PCs) 
between the device servers and the PLC controllers, 
as shown in Fig.~\ref{overview}.

\subsubsection{c) Web}

Web presentation of the KEK linac status 
would consume larger CPU and network resources than dedicated applications. 
The considerable increase of web accesses in recent years implies that 
we will need more computer resources in our control system. 
A more serious problem is that 
we will need more man-power to maintain  
both dedicated applications and web-based services. 
We are eager for 
some tools which would enable us automatically generate web contents.

\subsection{Transition of Control Architecture}

We are now ready to discuss the long-term transition of 
the basic architecture of the control system for large accelerators. 
Taking into account various accelerator control systems 
in the past 20 years, 
the transitions of control standards are given in Table~\ref{architecture} (upper).
Private expectations for the next decade (2000'es) 
are shown in Table~\ref{architecture} (lower). 

%
\vspace{-3 mm}
\small
\begin{table}[htb]
\begin{center}
\caption{Transition of large accelerator control.}
\begin{tabular}{|c||l|l|} \hline
          & 1980'es   & 1990'es    \\ \hline \hline
console   & CUI       & GUI on X   \\
          &(text-base)&(window-base) \\ \hline
base-machine     & mini-computer& Unix workstation  \\ 
(language)& FORTRAN   & C          \\ \hline
network   & dedicated network      & standard Ethernet \\
          & dedicated protocol & on TCP/IP \\ \hline
local-    & CAMAC     & VME with   \\
controller&           &   RT-OS    \\ \hline
\end{tabular}
\begin{tabular}{c} 
\\
\end{tabular}
\begin{tabular}{|c||l|} \hline
          & 2000'es (private expectation)  \\ \hline \hline
          & a) GUI on Windows-PC (for operation)  \\
console   & b) toolkit/environment (for study)   \\ 
          & c) Web/cell-phone (for announcement) \\ \hline
base-machine& Linux, or 64/128bit Unix   \\ 
(language)& Java, C++   \\ \hline
network   &  TCP/IP and CORBA  \\ 
          &  http (for web)    \\ \hline
local-    &  PLC with Ethernet (for simple I/O) \\
controller&  Linux box (for intelligent controller) \\ \hline
\end{tabular}
\label{architecture}
\end{center}
\end{table}
\normalsize
\vspace{-3 mm}

The present control system for the KEK linac, which started  in 1993, 
can be expressed as a typical standard of the 1990'es model. 
We conclude that 
our extensions (introducing subsystems in the past eight years)  
can be understood as an evolution toward the new standard of 2000'es.

%
%

%

%
%

\section{Acknowledgment}

The authors acknowledge Prof. A.~Enomoto and Prof. K.~Nakahara 
for kindly supervising our work. 
We thank I.~Abe for discussions on 
using Windows PCs with Unix workstations. 
The various improvements of PLC-based local controllers 
have been carried out by A.~Shirakawa. 
We also thank the KEK linac operators 
for cooperative and successful works to improve our control system.

%
%

\end{document}